\documentclass[aps,prc,twocolumn,superscriptaddress]{revtex4}

\usepackage{epsfig}
\begin{document}

{\bf Accepted for publication in Phys. Rev. C} \vspace{0.7cm}

\title{Giant dipole resonance width and the universality of the Critical Temperature included Fluctuation Model}

\author{Deepak Pandit}
\email[e-mail:]{deepak.pandit@vecc.gov.in}
\affiliation{Variable Energy Cyclotron Centre, 1/AF-Bidhannagar, Kolkata-700064, India}

\author{Srijit Bhattacharya}
\affiliation{Department of Physics, Barasat Govt. College, Barasat, N 24 Pgs, Kolkata - 700124, India }

\author{Balaram Dey}
\affiliation{Variable Energy Cyclotron Centre, 1/AF-Bidhannagar, Kolkata-700064, India}

\author{Debasish Mondal}
\affiliation{Variable Energy Cyclotron Centre, 1/AF-Bidhannagar, Kolkata-700064, India}

\author{S. Mukhopadhyay}
\affiliation{Variable Energy Cyclotron Centre, 1/AF-Bidhannagar, Kolkata-700064, India}

\author{Surajit Pal}
\affiliation{Variable Energy Cyclotron Centre, 1/AF-Bidhannagar, Kolkata-700064, India}

\author{A. De}
\affiliation{Department of Physics, Raniganj Girls' College, Raniganj-713358, India}

\author{S. R. Banerjee}
\email[e-mail:]{srb@vecc.gov.in}
\affiliation{Variable Energy Cyclotron Centre, 1/AF-Bidhannagar, Kolkata-700064, India}


\date{\today}

\begin{abstract}
The universality of the Critical Temperature included Fluctuation Model (CTFM) in explaining the evolution of the giant dipole resonance (GDR) width as a function of angular momentum is examined in the light of recent experimental data on $^{144}$Sm and $^{152}$Gd. We compare both the data sets with the phenomenological formula based on the CTFM and the thermal shape fluctuation model (pTSFM). The CTFM describes both the data sets reasonably well using the actual ground state GDR width ($\Gamma_0$) values, whereas, the pTSFM describes the $^{144}$Sm data well but is unable to explain the  $^{152}$Gd data using a single value of  $\Gamma_0$ for two excitation energies. These interesting results clearly indicate that the phenomenological CTFM can be used universally to describe the evolution of the GDR width with both angular momentum and temperature in the entire mass region. Moreover, it should provide new insights into the modification of the TSFM. 

\end{abstract}
\maketitle

\section{Introduction}

The study of collective motions in hot and fast rotating nuclei, especially Giant Dipole Resonance (GDR), provides a unique probe to explore the various kinds of structure (triaxial, prolate, oblate, spherical) that the nuclear system can assume at high temperature ($T$) and angular momentum ($J$) \cite {hara01, gaar92, Snov01}. This vibrational mode of nucleus is described as the out of phase oscillation between the protons and the neutrons. Over the years, intensive experimental studies of the GDR built on highly excited states of nuclei have shown that the GDR width increases with both T and J \cite{Kici02, Bra04, Rama01, Matt01, Kelly01, Dreb01, Baum01, Kmie01, Sri08}. 
This increase of the GDR width is described reasonably well within the theoretical Thermal Shape Fluctuation Model (TSFM). The TSFM is based on large amplitude thermal fluctuation of nuclear shape under the assumption that the time scale associated with thermal fluctuation is slow compared to GDR vibrations and the observed GDR strength function is the weighted average of all the shapes and orientations \cite{alh88, Orma01, Kus98}.
A systematic study of the thermal fluctuation model revealed the existence of a universal scaling law (pTSFM) for the apparent width of the GDR for all T, J and A \cite{Kus98}. Although pTSFM shows an increase of the GDR width with T, the model differs significantly from the experimental data at low temperatures ($T \leq$ 1.5 MeV) \cite{dipu12, supm11, heck03, cam03, dipu10}. 
In a recent work, a Critical Temperature included Fluctuation Model (CTFM) \cite{dipu12} was put forward incorporating an essential modification on the pTSFM, which was found to predict successfully the evolution of the GDR width of many different nuclei with temperature and angular momentum (at least up to T $\sim$ 3.0 MeV and J $\sim$ 60 $\hbar$). 
The CTFM emphasizes on a crucial point,  overlooked in the pTSFM \cite{Kus98}, that the GDR vibration itself induces a quadrupole moment even at $T$ = 0 MeV causing the nuclear shape to fluctuate. Therefore, when the giant dipole vibration having its own intrinsic fluctuation is used as a probe to view the thermal shape fluctuations, it is unlikely to feel the thermal fluctuations that are smaller than its own intrinsic fluctuation. As a result, the experimental GDR width should remain nearly constant at the ground state values up to a critical temperature ($T_c$) and the effect of thermal fluctuations should become evident only when they become larger than the intrinsic GDR fluctuations \cite{dipu12}. Moreover, the value of $\Gamma_0$ in CTFM is always adopted from the existing ground state GDR width systematics of nuclei, whereas, the pTSFM takes the ground state GDR width $\Gamma_0$ generally as 
a free adjustable parameter ($\sim$ 3.8 MeV for all nuclei) without any proper justification \cite{Kus98}.

A recent experimental investigation \cite{Ish13} populating the compound nucleus ($CN$) $^{144}$Sm, 
in the reactions $^{28}$Si + $^{116}$Cd at E($^{28}$Si) $\sim$ 125 and 140 MeV, demonstrates the evolution of the GDR width  with $J$ at low temperature ($T\leq$ 1.5 MeV). According to this recent experiment, the phenomenological parameterization pTSFM \cite{Kus98} predicts the angular momentum dependent GDR width successfully. In this regard, it is worthwhile to mention that previously a few experimental investigations \cite{ Sri08, Drc10, Bra04} on J-dependence of the GDR width have indicated the underestimated predictions of the pTSFM in comparison to the experimental data, especially in the higher angular momentum region. To this end, the whole scenario appears to be confusing. 
Thus, it appears to be interesting and tempting as well to exploit the above mentioned experimental data on $^{144}$Sm as a testing ground of the universality of our recently proposed CTFM.   

The evolution of the GDR width as a function of angular momentum has also been studied recently for $^{152}$Gd at two different excitation energies in the reactions $^{28}$Si + $^{124}$Sn at E($^{28}$Si) $\sim$ 149 and 185 MeV \cite{Drc10}. The authors found that two different values of $\Gamma_0$ (3.8 and 4.8 MeV) were required in the pTSFM to explain the experimental systematics at the two incident energies (149 and 185 MeV, respectively). They concluded that the contributions from both the inhomogeneous damping and the intrinsic collisional damping processes should be included. They proposed an empirical relation for the increase of the GDR width as a function of $T$ and $J$. However, they used the ground state value $\Gamma_0$ $\sim$ 3.8 MeV which is much smaller than the actual ground state value of $^{152}$Gd \cite{hara01, gaar92}. Therefore, it is relevant to test the CTFM for $^{152}$Gd for the two excitation energies.
The present work aims at examing the universality of CTFM by exploiting the data sets on $^{144}$Sm and $^{152}$Gd and at the same time
comparing the predictions of CTFM with those due to pTSFM.

\section{Data Analysis and Results}

In heavy ion fusion reactions, the $\gamma$-rays from the GDR-decay are emitted from various stages of decay cascade of the CN and thus, average values should be considered. The average values of $T$ and $J$ of the CN for GDR gamma decay should be different and less than those of the initial compound nucleus. On the other hand, it is also not proper to include each step in the CN decay chain for the averaging. 
Instead, for the purpose, one should incorporate only that part of the decay cascade which is contributing to the GDR $\gamma$-emission, 
thereby, setting a lower limit for the excitation energy in the CN decay cascade. This lower limit is estimated from the divided plot of the high energy $\gamma$-spectrum when the cutoff in the excitation energy only affects the $\gamma$-emission at very low energies but does not alter the GDR width.  
The procedure is discussed in detail in Ref  \cite{wie06, Sri08a}. Although an average temperature was estimated in Ref \cite{Ish13} for $^{144}$Sm, it was calculated considering all the decay steps thereby lowering $T_{avg}$. Moreover, no averaging was performed for the spin distribution and 
mean J was taken to be that of the initial compound nucleus which further reduced the average value of T. 
Therefore, $J_{avg}$ and $T_{avg}$ have been re-estimated applying the procedure discussed in Ref \cite{wie06, Sri08a} using a modified version of the statistical model code CASCADE \cite{pul77, Sri08a}. The average T was estimated as $T_{avg}$=[($\overline{E^*}$ - $\overline{E}_{rot}$ - E$_{GDR}- \Delta_p$)/$a(\overline{E^*})$]$^{1/2}$, where $a(\overline{E^*}$) is the energy dependent level density parameter and $\Delta_p$ is the pairing energy. $\overline{E}_{rot}$ was evaluated with $J_{avg}$ re-estimated using the above mentioned lower limit in E$^*$, while $\overline{E^*}$ was calculated by averaging E$^*$ with corresponding weights over the daughter nuclei in the CN decay cascade for the $\gamma$-emission in the GDR energy range 10-20 MeV. The initial J$_{CN}$ for each fold has been selected from the given J distribution in Ref \cite{Ish13}. The energy loss through half the thickness of the target was also included as suggested by the authors by populating the CN at 68 and 80 MeV corresponding to the incident energies of 125 and 140 MeV, respectively. The Ignatuyk-Reisdorf level density prescription \cite{Rei81, igna75} was adopted keeping the asymptotic level density parameter $\widetilde{a}$=A/8.5 MeV$^{-1}$. The re-estimated values of J$_{avg}$ and T$_{avg}$ are given in Table-\ref{data} and used in both CTFM and pTSFM for proper theoretical explanation.

The T dependence of the GDR width in CTFM by including the GDR induced fluctuation is given as \cite{dipu12} 
\begin{eqnarray}
\Gamma(T, J=0, A) =  \Gamma_0(A)  \;\;\;\;\;\; T \leq T_c \nonumber
\end{eqnarray}
\begin{equation}
\Gamma(T, J=0, A) = c\left(A\right) \ln\left(\frac{T}{T_c}\right) +  \Gamma_0(A) \;\;\;\;\;\; T > T_c \label{eqn1}
\end{equation}
where 
\begin{eqnarray}
T_c = 0.7 + 37.5/A \nonumber
\end{eqnarray}
\begin{eqnarray}
c(A) = 8.45 - A/50 \nonumber
\end{eqnarray}
and the J dependence of CTFM is given by the power law 
\begin{equation} 
\Gamma_{red} = \left[\frac{\Gamma_{exp}(T, J, A)}{\Gamma(T, J=0, A)}\right]^{\frac{T+3.3T_c}{7T_c}} = L\left(\xi\right) \label{eqn2}
\end{equation}
where L($\xi$) = 1 + 1.8/[1 + e$^{(1.3 - \xi)/0.2}$] and $\xi$ = J/A$^{5/6}$.

\begin{table}
\caption{\label{data} The $T_{avg}$ and $J_{avg}$ for the reactions 
$^{28}$Si + $^{116}$Cd at two beam energies of 125 and 140 MeV populating $^{144}$Sm.}  
\begin{center}		
\begin{tabular}{|c|c|c|c|c|c|c|c|}
\hline
E$_{Lab}$   &    Fold         &  $J_{CN}$ & $T_i$ &       $J_{avg}$      & $T_{avg}$ & E$_{GDR}$       & $\Gamma_{GDR}$ \\
(MeV)       &                 & ($\hbar$) & (MeV) &       ($\hbar$)      &    (MeV)  &    (MeV)        &    (MeV)        \\ \hline
 125        &   9 - 10        &    30.7   & 1.65  &           27         &    1.51   & 15.1 $\pm$ 0.3  &  8.2 $\pm$ 0.5   \\
 125        &  11 - 12        &    36.2   & 1.60  &           33         &    1.47   & 15.1 $\pm$ 0.2  &  8.2 $\pm$ 0.4   \\
 125        &  13 - 27        &    45.0   & 1.52  &           41         &    1.36   & 15.2 $\pm$ 0.2  &  8.4 $\pm$ 0.5   \\\hline
 140        &   7 - 8         &    25.9   & 1.89  &           23         &    1.67   & 15.1 $\pm$ 0.3  &  8.8 $\pm$ 0.5   \\
 140        &   9 - 10        &    32.4   & 1.84  &           29         &    1.60   & 15.2 $\pm$ 0.2  &  8.7 $\pm$ 0.5   \\
 140        &  11 - 12        &    38.9   & 1.77  &           35         &    1.56   & 15.1 $\pm$ 0.2  &  8.5 $\pm$ 0.4   \\
 140        &  13 - 14        &    45.1   & 1.72  &           41         &    1.49   & 15.0 $\pm$ 0.3  &  8.3 $\pm$ 0.4   \\
 140        &  15 - 16        &    50.9   & 1.65  &           46         &    1.42   & 15.2 $\pm$ 0.2  &  9.0 $\pm$ 0.5   \\
 140        &  17 - 27        &    59.2   & 1.57  &           54         &    1.32   & 15.5 $\pm$ 0.4  &  9.8 $\pm$ 0.6   \\ 
\hline
\end{tabular}
\end{center}		
\end{table} 

\begin{figure}
\begin{center}
\includegraphics[height=8.0 cm, width=7.0 cm]{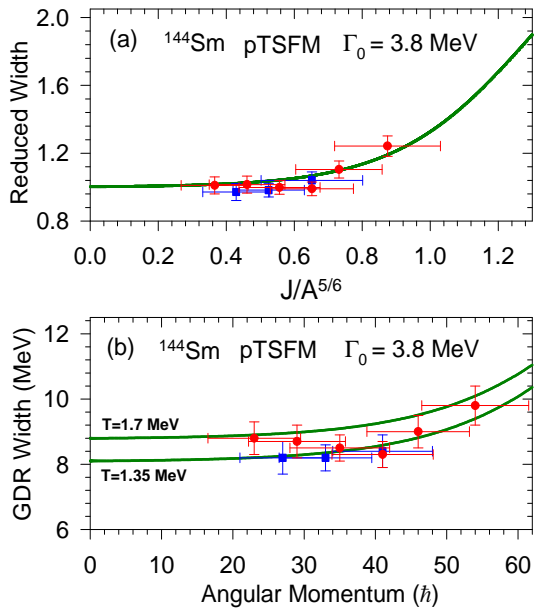}
\caption{\label{fig1} (color online) The filled squares and the filled circles represent the experimental data for the reactions $^{28}$Si + $^{116}$Cd at 125 and 140 MeV incident beam energies, respectively, populating $^{144}$Sm. (a) The reduced widths calculated using the pTSFM parameterization for the two beam energies are compared with the universal scaling function L($\xi$) (continuous line). (b) The experimental data are directly compared with the predictions of pTSFM (continuous lines) as a function of angular momentum for the two extreme temperatures involving the experimental data.}
\end{center}
\end{figure}

\begin{figure}
\begin{center}
\includegraphics[height=12.0 cm, width=7.5 cm]{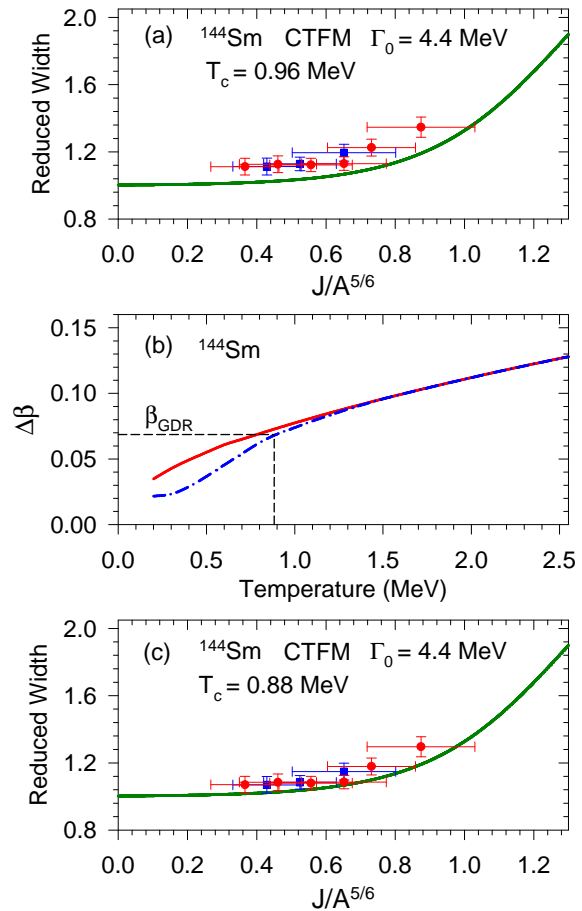}
\caption{\label{fig2} (color online) The symbols are same as those in Fig.\ref{fig1}. (a) The reduced widths calculated using the CTFM formalism for the two beam energies are compared with the universal scaling function L($\xi$) (continuous line). (b) The variation of $\Delta\beta$ as function of T for $^{144}$Sm with shell effect(dotted-dashed line) and without shell effect (continuous line). The corresponding $\beta_{GDR}$ (dashed line) is compared with $\Delta\beta$ to extract the critical temperature ($T_c$). (c) The reduced widths calculated within the CTFM formalism using the extracted value of $T_c$ are compared with the universal scaling function L($\xi$) (continuous line).}
\end{center}
\end{figure}

\subsection{$^{144}$Sm Nucleus}
\begin{figure}
\begin{center}
\includegraphics[height=8.0 cm, width=7.0 cm]{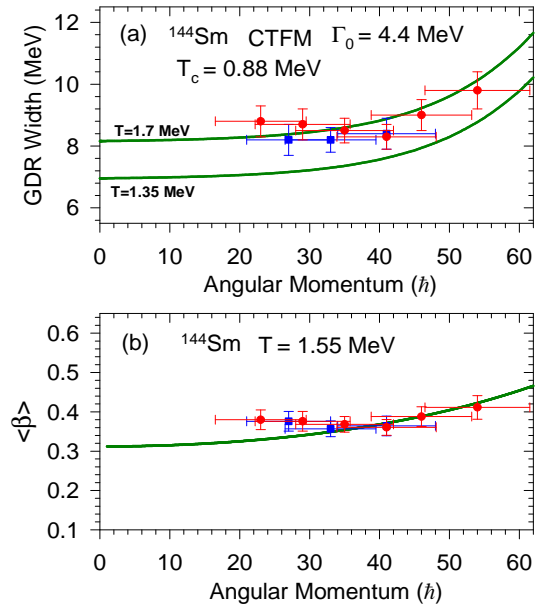}
\caption{\label{fig3} (color online) The symbols are same as those in Fig.\ref{fig1}. (a) The experimental data are directly compared with the predictions of CTFM (continuous lines) as a function of angular momentum  for the two extreme temperatures involving the experimental data. (b) The empirical deformations extracted from the experimental GDR widths using the universal correlation are compared with the TSFM as a function of angular momentum (continuous line).}
\end{center}
\end{figure}

The reduced GDR widths ($\Gamma_{red}$) with the reduced parameter $J/A^{5/6}$ for different temperatures of $^{144}$Sm applying the pTSFM formalism are shown in Fig.\ref{fig1}(a). The parameter $\Gamma_0$ was taken as 3.8 MeV for the calculations. As could be seen, the experimental data are in reasonable agreement with the universal scaling function, as concluded in Ref \cite{Ish13}. We also compare the experimental data directly with the predictions of pTSFM as a function angular momentum (Fig.\ref{fig1}(b)). The pTSFM calculation was performed for the two extreme temperatures involving the entire T domain of the experimental data (T $\sim$ 1.35 - 1.7 MeV). Interestingly, in this case too, the pTSFM predictions are in good agreement with the experimental data. In order to compare the data with CTFM, the ground state GDR width ($\Gamma_0$) of $^{144}$Sm was estimated considering the small ground state deformation ($\beta$ = 0.0874) \cite{ram01} and spreading width parameterization $\Gamma$$_s$=0.05E$^{1.6}_{GDR}$ \cite{jun08} for each Lorentzian. This new empirical formula for the spreading width has been derived by separating the deformation induced widening from the spreading effect and requiring the integrated Lorentzian curves to fulfill the dipole sum rule. The ground state value was estimated to be 4.4 MeV which is consistent with the experimentally measured value of 4.37 $\pm$ 0.15 MeV  \cite{Car74}. The reduced GDR widths with the reduced parameter $J/A^{5/6}$ for CTFM are shown in Fig.\ref{fig2}(a). It is observed that CTFM overpredicts the experimental data when $T_c$ is obtained from the systematics ($T_c$ = 0.96 MeV). Therefore, we estimated the critical temperature directly by comparing the intrinsic GDR fluctuation ($\beta_{GDR}$) due to induced quadrupole moment with the variance of the deformation ($\Delta\beta$) due to thermal fluctuations as shown in Fig.\ref{fig2}(b). The $\Delta\beta$ was calculated using the Boltzmann probability $e^{-F(\beta, \gamma)/T}$ with the volume element $\beta^{4} sin(3\gamma) d\beta d\gamma$, according to the formalism described in Ref \cite{Dipu2} while $\beta_{GDR}$ was estimated from the systematic $\beta_{GDR}$ = 0.04 + 4.13/A \cite{dipu12, Dipu13}. 
The $\Delta\beta$ values were calculated with and without considering shell effect and is represented by dotted-dashed and continuous lines, respectively, 
in Fig.\ref{fig2}(b). As could be seen, $\Delta\beta$ and $\beta_{GDR}$ are equal at T = 0.88 MeV (including shell effect) and T = 0.8 MeV (without shell effect). The inclusion of shell effect increases the critical temperature making it closer to the value predicted by the systematics. Using the value of $T_c$ = 0.88 MeV, the reduced widths within the CTFM formalism were calculated and compared with the universal scaling law. It is interesting to note that the experimental data now match well with the CTFM predictions. The data have also been compared directly with the predictions of CTFM as a function of angular momentum. It is evident from  Fig.\ref{fig3}(a) that CTFM represents the experimental data reasonably well using the calculated $T_c$ and the actual ground state value of the GDR width. Therefore, it can be inferred that the predictions of both pTSFM and CTFM are consistent with the experimental findings in the case of  $^{144}$Sm.
Recently,  a universal correlation between the experimental GDR width and the average deformation $\left\langle\beta\right\rangle$ of the nucleus 
at finite excitation have been proposed by including the deformation induced by the GDR motion. In order to verify the correlation for $^{144}$Sm, the empirical deformation has been estimated from the experimental GDR width applying the correlation given in Ref \cite{Dipu13}. The $\left\langle\beta\right\rangle$ values extracted from the experimental data are directly compared with the TSFM calculation (continuous line) in Fig.\ref{fig3}(b). As could be seen, the empirical deformations extracted from the experimental data are in good agreement with the TSFM calculation for $^{144}$Sm. This excellent match between experimental data and the TSFM clearly points toward the fact that this universal correlation between $\left\langle\beta\right\rangle$ and GDR width provides a direct experimental probe to assess the nuclear deformation at finite temperature and angular momentum. 

\begin{figure}
\begin{center}
\includegraphics[height=8.0 cm, width=7.0 cm]{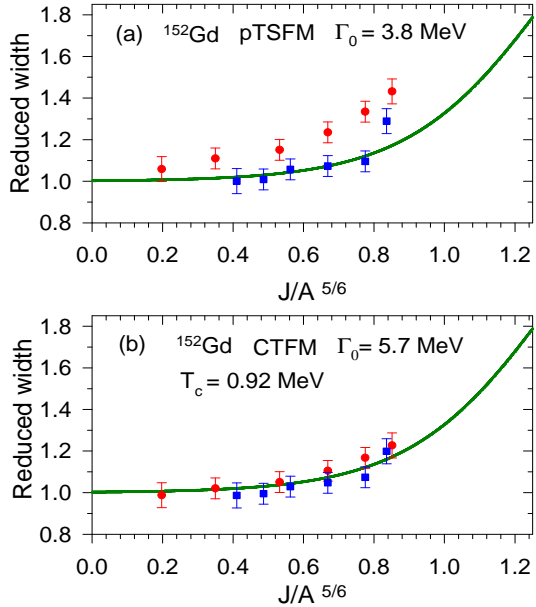}
\caption{\label{fig4} (color online) The filled squares and the filled circles represent the experimental data for the reactions $^{28}$Si + $^{124}$Sn at 149 and 185 MeV incident beam energies, respectively, populating $^{152}$Gd. a) The reduced widths calculated using the pTSFM parameterization for the two beam energies are compared with the universal scaling function L($\xi$) (continuous line). (b) The reduced widths calculated using the CTFM parameterization are compared with the universal scaling function L($\xi$) (continuous line). }
\end{center}
\end{figure}
\begin{figure}
\begin{center}
\includegraphics[height=4.0 cm, width=7.0 cm]{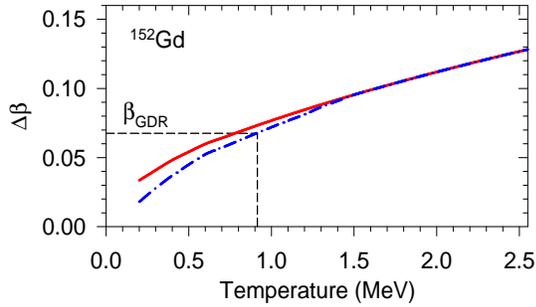}
\caption{\label{fig5} (color online) The variation of $\Delta\beta$ as function of T for $^{152}$Gd with shell effect (dotted-dashed line) and without shell effect (continuous line). The corresponding $\beta_{GDR}$ (dashed line) is compared with $\Delta\beta$ to extract the critical temperature ($T_c$). }
\end{center}
\end{figure}
\begin{figure}
\begin{center}
\includegraphics[height=8.0 cm, width=7.0 cm]{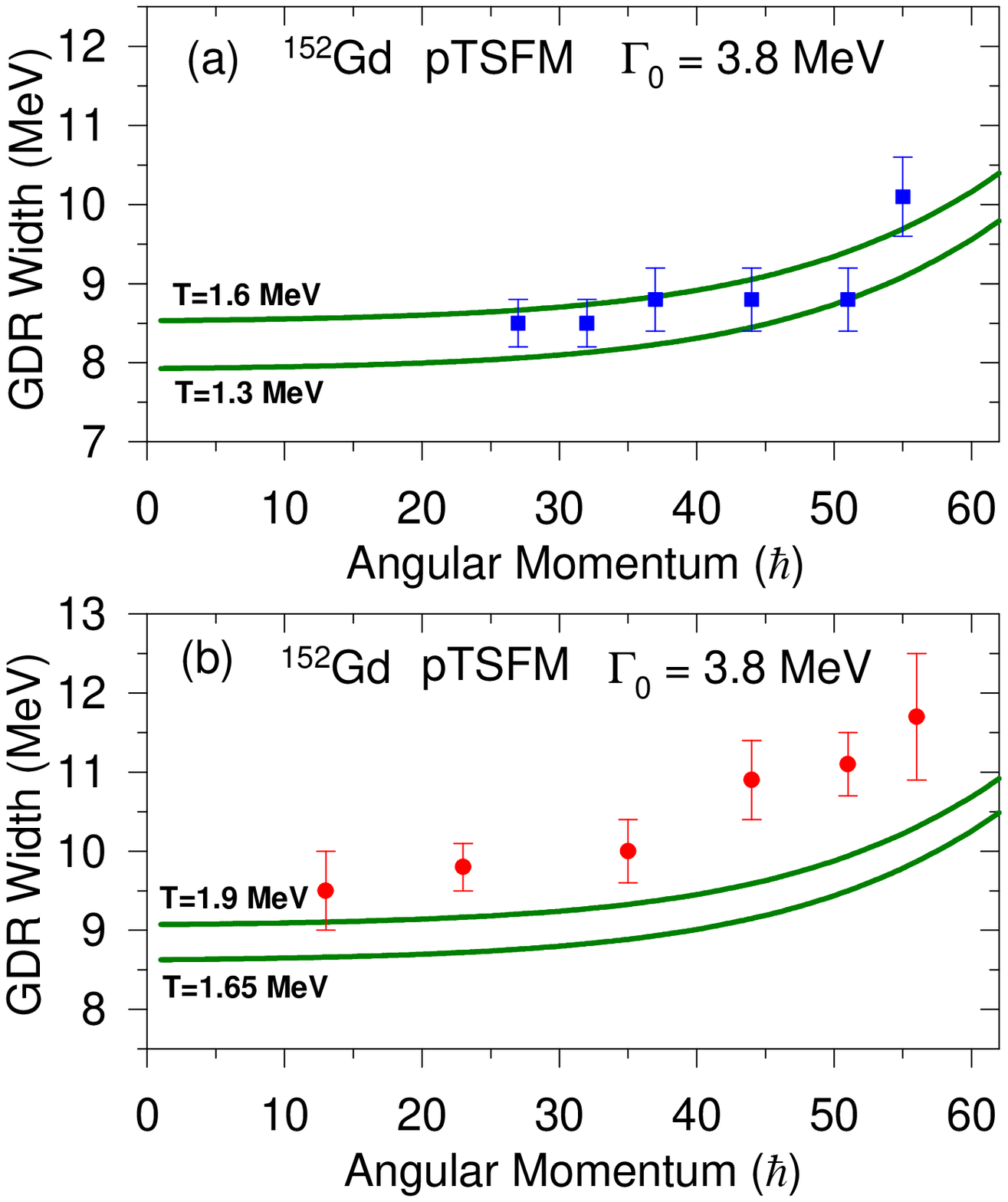}
\caption{\label{fig6} (color online) (a) The experimental data (filled squares) for the reaction $^{28}$Si + $^{124}$Sn at 149 MeV are directly compared with the predictions of pTSFM (continuous lines) as a function of angular momentum for the two extreme temperatures involving the experimental data. (b) Same as above but for the reaction $^{28}$Si + $^{124}$Sn at 185 MeV.}
\end{center}
\end{figure}
\begin{figure}
\begin{center}
\includegraphics[height=8.0 cm, width=7.0 cm]{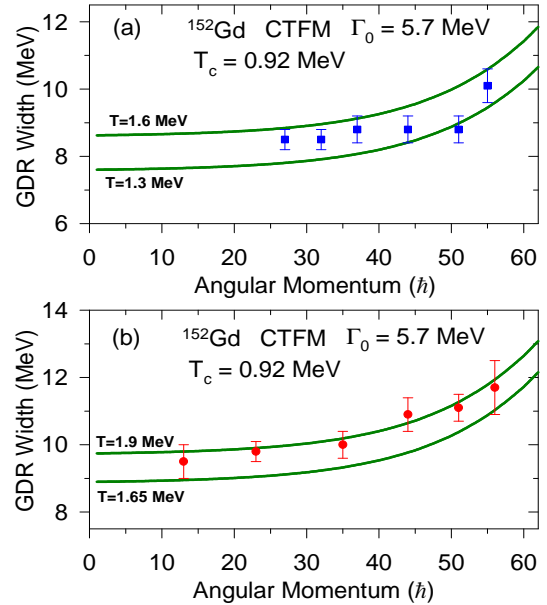}
\caption{\label{fig7} (color online) (a) The experimental data (filled squares) for the reaction $^{28}$Si + $^{124}$Sn at 149 MeV are directly compared with the predictions of CTFM (continuous lines) as a function of angular momentum for the two extreme temperatures involving the experimental data. (b) Same as above but for the reaction $^{28}$Si + $^{124}$Sn at 185 MeV.}
\end{center}
\end{figure}

\subsection{$^{152}$Gd Nucleus}
The reduced GDR widths are plotted against the reduced parameter $J/A^{5/6}$ at two different excitation energies in the reactions $^{28}$Si + $^{124}$Sn at E($^{28}$Si) $\sim$ 149 and 185 MeV \cite{Drc10} according to the pTSFM formalism and are shown in Fig.\ref{fig4}(a).
From Fig.\ref{fig4}(a) it could be seen that the pTSFM gives good description of the data at 149 MeV but is unable to represent the experimental data at 185 MeV using $\Gamma_0$ = 3.8 MeV, as concluded in Ref \cite{Drc10}.
The ground state GDR width as well as the critical temperature for $^{152}$Gd were estimated in order to compare the data with CTFM. Interestingly, $\Delta\beta$ and $\beta_{GDR}$ are equal at T = 0.92 MeV (including shell effect) (Fig.\ref{fig5}) which is similar to the value predicted by the systematics ($T_c$ = 0.94 MeV). The value of   $\Gamma_0$ was estimated as 5.7 MeV considering the large ground state deformation of $\beta$ = 0.206 \cite{ram01} and spreading width parameterization $\Gamma$$_s$=0.05E$^{1.6}_{GDR}$ \cite{jun08} for each Lorentzian. Using these values of $\Gamma_0$ and $T_c$, the reduced GDR widths were calculated and compared with the universal scaling function. It is very interesting to note that the CTFM represents the experimental data remarkably well for both the excitation energies using the actual value (5.7 MeV) of the GDR ground state width (Fig.\ref{fig4}(b)). The experimental data have also been compared directly with the predictions of pTSFM and CTFM as a function angular momentum (Fig.\ref{fig6} and Fig.\ref{fig7}). 
The discrepancy of pTSFM is more evident in this representation as it explains the data at 149 MeV but underpredicts the data at 185 MeV using  $\Gamma_0$ = 3.8 MeV (Fig.\ref{fig6}). The CTFM, on the other hand, successfully describes the experimental data for both the incident energies 149 MeV and 185 MeV covering the temperature domain T = 1.3 - 1.6 MeV and T=1.65-1.9 MeV, respectively (Fig.\ref{fig7}).  
As it appears, pTSFM represents the correct trend of J-dependence of the GDR width but is unable to explain the $^{152}$Gd data because it does not represent the correct description of the evolution of the GDR width with T. However, it needs to be mentioned that the pTSFM was formulated at a time when there was no experimental data for GDR widths at low temperatures. The empirical deformations extracted from the experimental GDR widths for $^{152}$Gd have already been compared with the TSFM in Ref \cite{ Dipu13}. Interestingly, in this case too, the data and the TSFM were found to be in good agreement with each other. Thus, the good match of the CTFM with the experimental data for both $^{144}$Sm and $^{152}$Gd clearly points toward the fact that the phenomenological CTFM can be used efficiently to describe the evolution of the GDR width with both T and J in the entire mass region.

\section{Summary and Conclusions}
In conclusion, the CTFM gives good description of the J-dependence of GDR width systematics, for both $^{144}$Sm and $^{152}$Gd studied at two different excitation energies, using the actual ground state GDR width values for both the nuclei. On the other hand, the pTSFM based on the thermal shape fluctuation model explains the data for $^{144}$Sm but cannot represent the $^{152}$Gd data for the two excitation energies using a single value of $\Gamma_0$. The good description of the CTFM for both $^{144}$Sm and $^{152}$Gd as well as the validity of the universal correlation between the average deformation of the nucleus and the TSFM should provide new insights into the modification of the TSFM by including the GDR induced quadrupole moment to explain the GDR width systematics at low T.

\end{document}